\documentclass[proof]{WileyASNA-v1}

\articletype{Article Type}%
\usepackage{epsfig,amssymb,amsmath,rotating,lineno,graphicx}
\received{xxx 2025}
\revised{xxx 2026}
\accepted{xxx 2026}

\begin{document}

\title{Study of UV line and continuum variabilities in the Broadline Seyfert 1 Galaxy ESO 141$-$G55}

\author[1]{Mayukh Pahari*}

\author[2]{Veda Samhita}

\author[3]{Harikumar N.}

\author[1]{Anurag Baruah}

\author[1]{Vivek Shinde}

\address[1]{\orgdiv{Department of Physics}, \orgname{IIT Hyderabad}, \orgaddress{\state{Kandi, Sangareddy, Telangana}, \country{India}}}

\address[2]{\orgdiv{Department of Electronics and Communications Engineering}, \orgname{Vasavi College of Engineering, Hyderabad}, \orgaddress{\state{Telangana}, \country{India}}}

\address[3]{\orgdiv{Physics and Astronomy Department}, \orgname{NIT Rourkela}, \orgaddress{\state{Odisha}, \country{India}}}

\corres{*Mayukh Pahari; \email{mayukh@phy.iith.ac.in}}

\presentaddress{PH209, Department of Physics, IIT Hyderabad, Kandi, Sangareddy, Telangana, India}

\newcommand{\changes}[1]{{\color{red} #1}}

\abstract{We present the results from a 3-year-long Ultraviolet monitoring campaign of the broad line Seyfert 1 galaxy ESO 141$-$G55 using International Ultraviolet Explorer (IUE). By modelling all individual, extinction-corrected UV spectra in 1150$-$1978~\AA~ and 1850-3348~\AA~ wavelength range, we have observed a significant variability in both UV continuum and line fluxes. Variabilities due to ionised UV lines like Si\textsc{iv}, C\textsc{iv} and He\textsc{ii} are delayed with respect to the UV continuum by 2.92$^{+0.54}_{-0.61}$, 4.41$^{+0.44}_{-0.54}$, 4.11$^{+0.35}_{-0.81}$ days, respectively. At a distance of $\sim$0.004c, an outer accretion disc can be a possible site for the origin of UV lines. }


\maketitle

%\linenumbers

\section{Introduction}\label{sec1}

Seyfert 1 galaxies are a subclass of Active galactic nuclei (AGN) characterised by bright, point-like nuclei and broad permitted emission lines in their optical spectra, indicating the presence of high-velocity gas close to the central supermassive black hole \citep{Khachikian1974, Osterbrock1981}. These systems exhibit significant ultraviolet (UV) emission, which originates from the innermost regions of the accretion disk \citep{Shakura1973}. The UV continuum in Seyfert 1s often shows variability on timescales ranging from days to months, which is thought to be driven by instabilities or changes in the accretion flow and is frequently accompanied by corresponding changes in broad emission lines from broad line regions (BLRs) such as Ly$\alpha$ $\lambda$1216, C \textsc{iv} $\lambda$1549, and He \textsc{ii} $\lambda$1640 \citep{Koratkar1991, Clavel1991}. 

One of the most effective tools for probing the inner disk, BLR structure of AGN is reverberation mapping (RM) \citep{Blandford1982, Peterson1993}, particularly in UV. RM exploits the intrinsic variability of the continuum source - often the accretion disk and measures the time delays between fluctuations in the continuum and the corresponding response in the broad emission lines produced by photoionized gas in the BLR. For measurements of time lag, we apply classical cross$-$correlation methods \citep{1988ApJ...333..646E, Peterson1993} which uses randomization techniques like flux randomization and random subset selections. This lag represents the mean light travel time from the accretion disk to the BLR, thereby providing an estimate of the BLR’s average distance from the central source. The primary objective of this work is to investigate the location of BLR through lag measurements, particularly, distinguishing the boundary between the BLR and the outer accretion disk using UV reverberation measurements. An accretion disk surrounds the supermassive black hole (SMBH), and the optical and ultraviolet (UV) emission seen in the spectral energy distribution (SED) of an active galactic nucleus (AGN) is believed to arise from the disk’s thermal radiation. Early models typically assumed a temperature profile based on the classical prescription by \citet{Shakura1973}. In such a scenario, high-energy radiation striking the disk enhances its intrinsic thermal emission, resulting in a wavelength-dependent time delay, denoted by $\tau$, between the incoming high-energy photons and the reprocessed UV/optical emission. This delay follows the relation: $\tau \propto (M^2 \dot{m}_E)^{1/3} \lambda^{\beta}$, where $\beta = 4/3$, $M$ is the black hole mass, and $\dot{m}_E$ is the accretion rate expressed in Eddington units. Earlier investigations \citep{Collier1999,Cackett2007} supported the case where $\beta = 4/3$ using optical observations. Recent multiwavelength reverberation studies, such as \citet{2016ApJ...821...56F}, have shown that the measured UV-optical interband lags in NGC 5548 are a factor of about three larger than predicted by standard thin-disk models, indicating the presence of an “oversized” accretion disk. \citet{2022ApJ...940...20G} further confirmed this discrepancy using high-cadence UV-optical monitoring of several Seyfert 1 galaxies, suggesting that temperature fluctuations and diffuse continuum emission could contribute to the observed lag excess. More recently, \citet{2023ApJ...958..195C} reported similar results from the AGN STORM 2 campaign on Mrk 817, reinforcing the oversized-disk problem and highlighting the need to refine current accretion disk and reprocessing models.

On the other hand, located farther out from the accretion disk, at a distance of approximately 0.001-0.1 parsecs, typical for Seyferts, is the BLR, which is responsible for producing emission lines. The UV/optical continuum emitted by the accretion disk photoionizes the gas clouds within the BLR. Since these clouds reside deep within the gravitational potential of the black hole, their resulting emission lines are broadened to velocities of several thousand kilometres per second. Therefore, the site of the reprocessing region is difficult to probe given the complexity of the central region.

ESO 141--G55 is a nearby (\(z = 0.0371 \pm 0.0003\); \citealt{2004NuPhS.132..185G, vaucouleurs1991third}) broad-line Seyfert 1 galaxy (BLS1). It was initially identified as an X-ray-bright AGN by \citet{Elvis1978} in the Ariel 5 survey and confirmed as a Seyfert 1 through optical spectroscopy by \citet{Ward1978}. The galaxy exhibits a blue stellar nucleus and prominent optical and UV emission lines \citep{1978ApJ...223..788W}, along with significant variability in both the UV continuum and emission lines over timescales of months \citep{1981IAUC.3560....1G, 1985ApJ...297..151C}. Archival IUE spectra of ESO 141--G55 have shown strong intrinsic UV variability. For instance, the 140 nm flux doubled between June and October 1980, increasing from \(0.8 \times 10^{-15}\) to \(1.6 \times 10^{-15}\ \text{J m}^{-2} \text{s}^{-1} \text{nm}^{-1}\) \citep{1981IAUC.3560....1G}. In the X-ray regime, observations with \texttt{EXOSAT} revealed a flat 0.1--10 keV power-law slope with \(\Gamma = 1.52^{+0.03}_{-0.06}\) \citep{Turner1989}, while subsequent measurements with \texttt{Einstein} and \texttt{ROSAT} showed a steeper photon index, with \texttt{ROSAT} finding \(\Gamma = 2.39 \pm 0.09\), suggesting a soft X-ray excess \citep{Turner1991, Turner1993}. More recent observations using \texttt{XMM-Newton} revealed a complex X-ray spectrum that includes a smooth soft excess, a broad Fe K$\alpha$ line, and a pronounced Compton reflection hump, but no significant warm absorber features - leading to its classification as a bare AGN \citep{porquet2024revealing}. This bare-line-of-sight nature makes ESO 141-G55 an ideal target for reverberation studies, as it offers a clear, relatively unobscured view into the inner accretion disk and BLR.

In this paper, the ultraviolet properties of ESO 141--G55 have been explored using data from the International Ultraviolet Explorer (IUE), which observed the source in both the short and long wavelength ranges, covering multiple epochs between 1978 and 1982. The observations and data reduction methods are presented in Section \ref{sec2}. Section \ref{sec3} outlines the results obtained from the spectral analysis. In Section \ref{sec4}, we present the time lag analysis between the UV continuum and prominent emission lines. Section \ref{sec5}, discusses the origin of ultraviolet line emissions and investigates the structure and scale of the broad-line region. Finally, concluding remarks about this work are presented in Section \ref{sec6}.

\begin{figure*}
    \centering
    \begin{minipage}{0.48\textwidth}
        \includegraphics[width=\columnwidth]{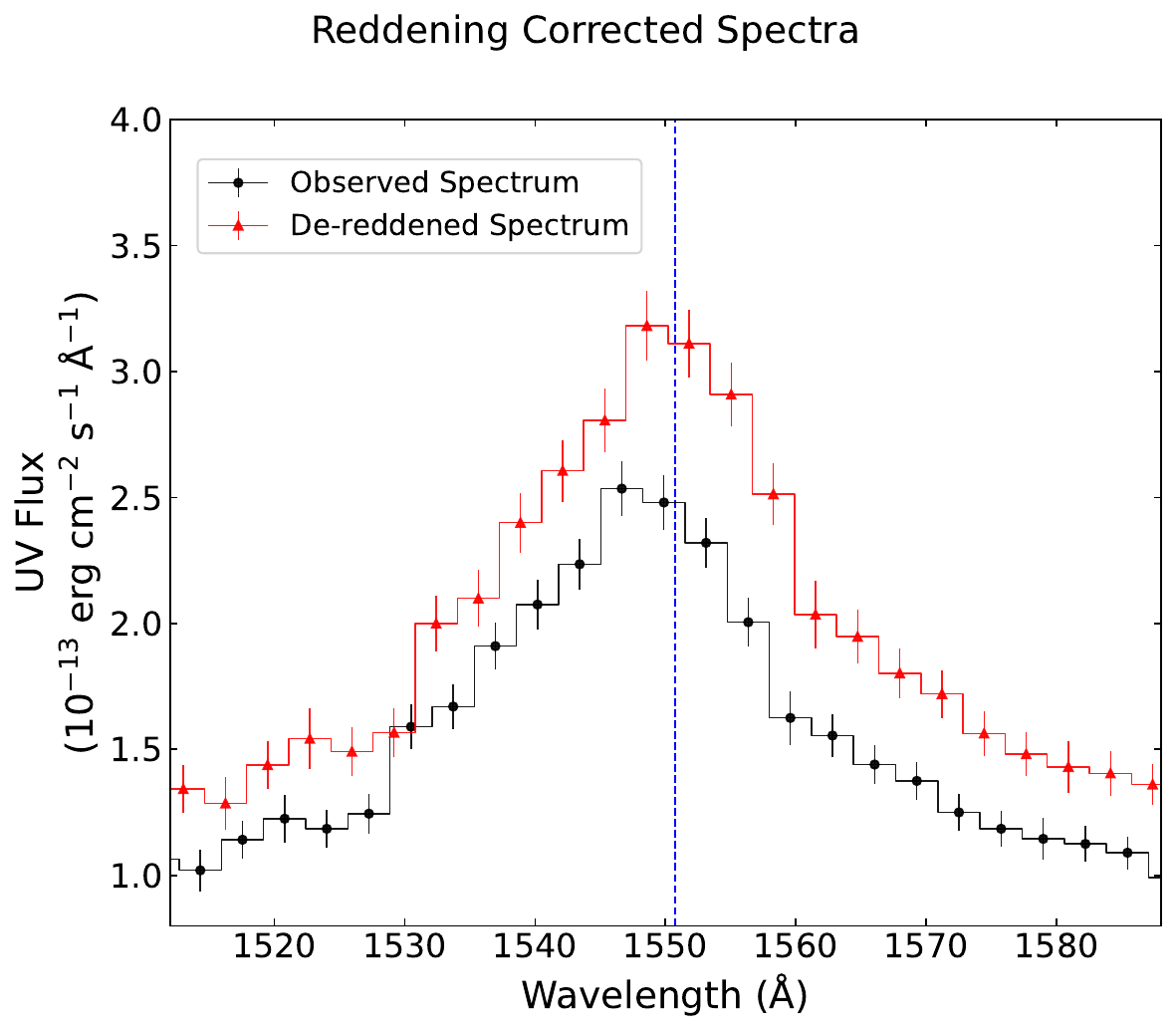}
        \caption{Reddening-corrected UV spectra showing the observed (black) and de-reddened (red) flux distributions. There is significant increase in flux levels after the correction. The vertical blue dashed line marks the offset-corrected C\textsc{iv} emission feature at 1550.77~\AA.}
        \label{fig:deredden}
    \end{minipage}
    \hfill
    \begin{minipage}{0.48\textwidth}
        \includegraphics[width=\columnwidth]{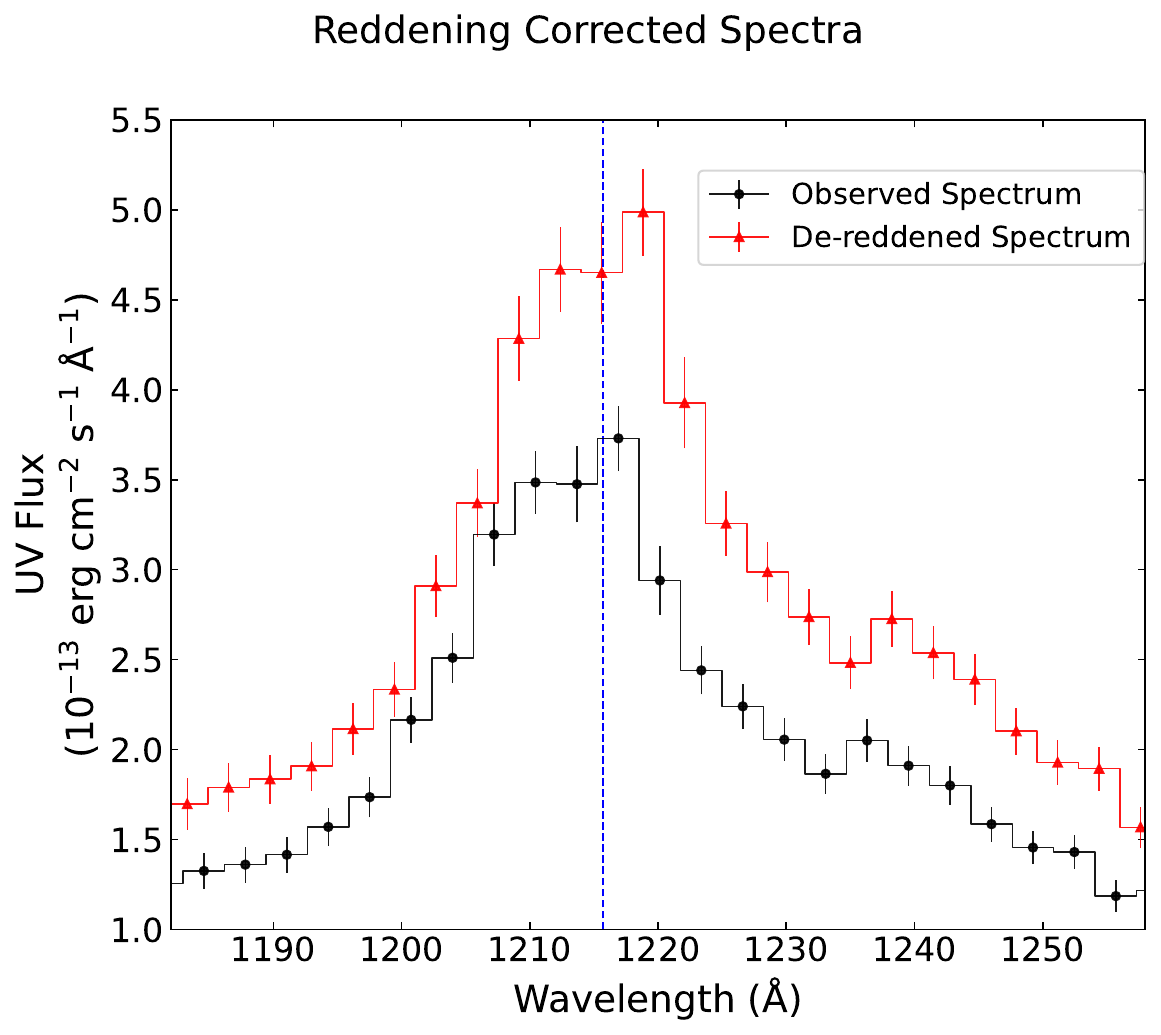}
        \caption{Reddening-corrected UV spectra showing the observed (black) and de-reddened (red) flux distributions. There is significant increase in flux levels after the correction. The vertical blue dashed line marks the offset-corrected Ly$\alpha$ emission feature at 1215.67~\AA.}
        \label{fig:deredden2}
    \end{minipage}
\end{figure*}

\section{OBSERVATIONS AND DATA REDUCTION}\label{sec2}

Between June 14, 1979 and March 15, 1991, the International Ultraviolet Explorer (IUE) performed 53 low-resolution short and long-wavelength spectral observations in the wavelength range of 1150-1978~\AA~and 1850-3348~\AA. The analysis procedure follows the methodology described in \citet{Wanders1997}. Raw IUE images are processed using the TOMSIPS \citep{Ayres1993} and NEWSIPS \citep{Nichols1993} data reduction pipelines. In this study, we adopt the NEWSIPS-reduced spectra, as recommended by \citet{Wanders1997}. The TOMSIPS reduction suffers from a nonlinear wavelength calibration error due to uncorrected long-term drifts in the wavelength scale. In contrast, the NEWSIPS pipeline provides spectra that agree well with HST data, without requiring additional corr- and long-wavelength ranges, spanningng the NEWSIPS pipeline, a small wavelength shift of approximately 1-2~\AA{} is introduced due to pointing errors associated with the large aperture \citep{Wanders1997}. To correct for this, an offset compensation is applied to align the sharp C\textsc{iv} line feature across all spectra to a common average wavelength. The calculated interstellar reddening is significant with E(B-V) = 0.111. The reddening correction is applied to each spectrum following \citet{Calzetti99}:
\begin{equation}
    F_{obs}(\lambda) = F_{actual}(\lambda) \times 10^{-0.4\times k(\lambda)\times E(B-V)}
\end{equation}
where $k(\lambda)= - 2.156 + \frac{1.509}{\lambda} - \frac{0.198}{\lambda^2} + \frac{0.011}{\lambda^3}$. The de-reddened spectra is shown in Figure \ref{fig:deredden} and Figure \ref{fig:deredden2}. The vertical blue lines indicate the offset corrected C\textsc{iv} emission at 1550.77~\AA~and Ly$\alpha$ emission at 1215.67~\AA~respectively. The redshift observed from the C\textsc{iv} average peak is consistent with the spectroscopic red-shift of z = 0.0371 of \citet{vaucouleurs1991third}.

Table \ref{tab1} lists the IUE observations of ESO 141--G55. Out of the 53 total observations, 8 do not have any prominent emission features, which are not suitable for this work. Therefore, we use 46 observations for further analysis.

\begin{table*}[htbp]
\centering
\caption{IUE data of ESO 141-G55.\label{tab1}}
\begin{tabular}{lcc|lcc}
\hline
Data ID & Observation Start Time & Exp Time (s) & Data ID & Observation Start Time & Exp Time (s)\\
 & (YYYY-MM-DD HH:MM:SS) &  &  & (YYYY-MM-DD HH:MM:SS) & \\
\hline
SWP05514 & 1979-06-14 11:14:03 & 3600 & SWP14399 & 1981-07-05 01:49:39 & 3000 \\
SWP09200 & 1980-06-05 20:56:49 & 3180 & SWP15256 & 1981-10-13 21:39:00 & 480 \\
SWP10269 & 1980-10-02 18:00:40 & 3600 & SWP15468 & 1981-11-09 16:04:19 & 3600 \\
SWP10270 & 1980-10-02 19:58:00 & 4800 & SWP16428 & 1982-02-26 03:06:21 & 3000 \\
SWP10298 & 1980-10-07 07:54:51 & 3000 & SWP16560 & 1982-03-18 00:37:23 & 3180 \\
SWP10643 & 1980-11-20 17:51:44 & 3000 & SWP16788 & 1982-04-18 17:55:30 & 3900 \\
SWP10644 & 1980-11-20 19:08:49 & 2280 & SWP17622 & 1982-08-07 20:08:21 & 3600 \\
SWP10741 & 1980-12-03 12:45:30 & 3600 & SWP18095 & 1982-09-24 15:23:07 & 1680 \\
SWP13453 & 1981-03-10 12:39:11 & 9540 & SWP18241 & 1982-10-08 12:30:45 & 4500 \\
SWP13454 & 1981-03-10 17:45:08 & 5160 & SWP18383 & 1982-10-24 12:47:06 & 3600 \\
SWP13563 & 1981-03-24 04:31:22 & 2700 & SWP18455 & 1982-11-02 12:56:40 & 3600 \\
SWP13617 & 1981-03-31 00:36:55 & 3000 & SWP18507 & 1982-11-08 17:00:23 & 3000 \\
SWP14275 & 1981-06-18 03:23:14 & 3600 & & & \\
\hline
LWR04779 & 1979-06-14 12:19:20 & 3600 & LWR10890 & 1981-06-18 04:27:05 & 4500 \\
LWR08940 & 1980-10-02 16:55:25 & 3600 & LWR11010 & 1981-07-05 02:44:15 & 3780 \\
LWR08941 & 1980-10-02 19:07:00 & 2700 & LWR11952 & 1981-11-09 17:08:27 & 3600 \\
LWR08962 & 1980-10-07 06:47:38 & 2400 & LWR13043 & 1982-04-19 16:14:00 & 5700 \\
LWR09352 & 1980-11-20 16:56:29 & 3000 & LWP01631 & 1982-08-07 19:14:48 & 3000 \\
LWR09425 & 1980-12-03 11:40:10 & 3600 & LWP01632 & 1982-08-07 21:12:03 & 3000 \\
LWR10116 & 1981-03-10 15:23:03 & 8280 & LWP01708 & 1982-11-08 16:06:49 & 3000 \\ 
LWR10198 & 1981-03-24 05:20:49 & 3300 & & & \\
\hline
\end{tabular}
\end{table*}

\section{UV SPECTRAL ANALYSIS AND RESULTS}\label{sec3}

\begin{figure}
    \centering
    \includegraphics[width=\columnwidth]{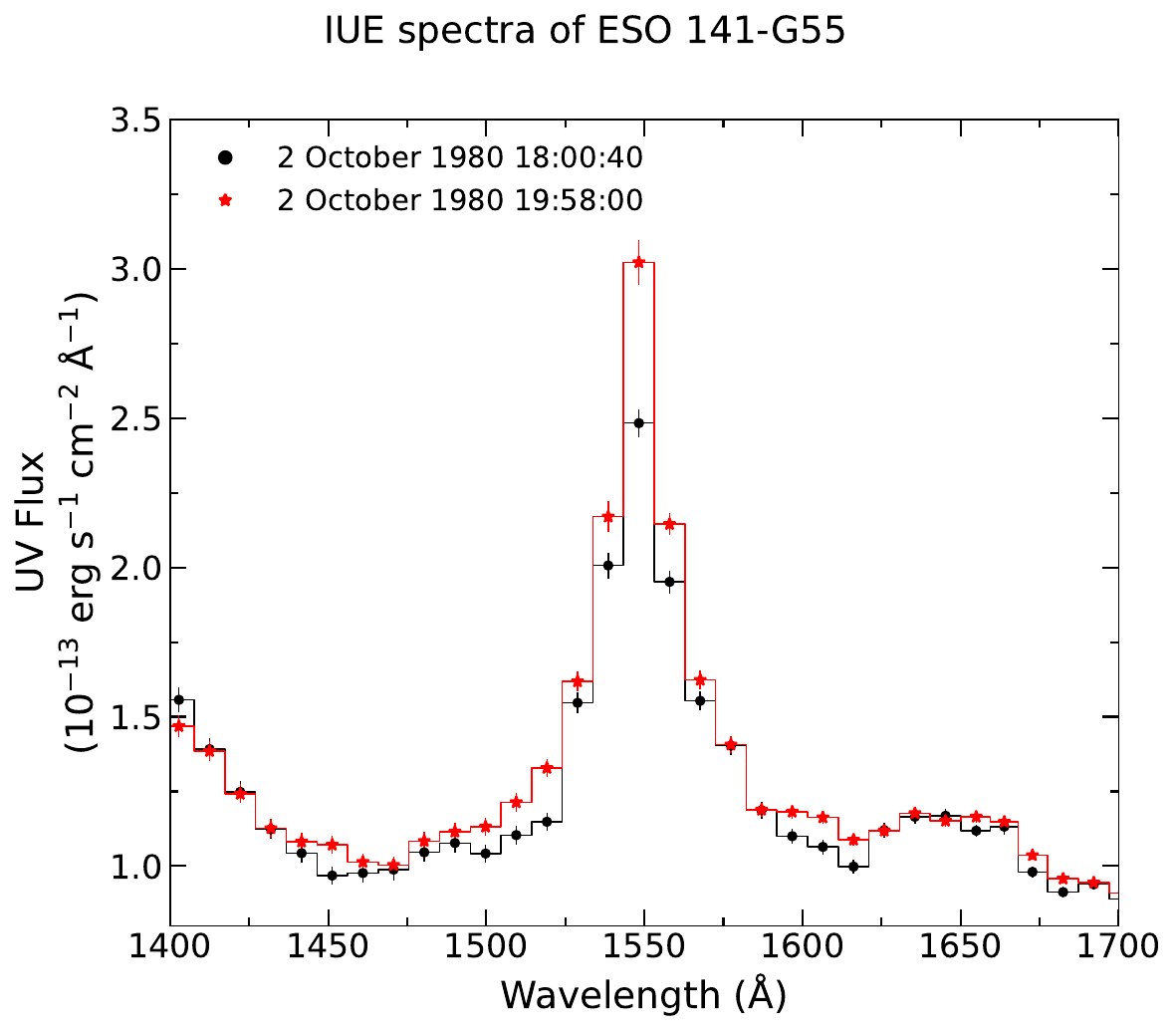}
    \caption{UV variability of ESO 141-G55 as observed on 5 June 1980 at 20:56:49 (circles) and 2 October 1980 at 19:58:00 (stars). IUE spectra in the wavelength range of 1550--1670~\AA~ show a substantial change in the flux level.}
    \label{Fig1}
\end{figure}

The IUE ultraviolet spectra exhibit significant variability in both line and continuum flux levels. An example of such variations is presented in Figure \ref{Fig1}, where the flux observed on 5 June 1980 at 20:56:49 is notably different than that recorded on 2 October 1980 at 19:58:00.

\begin{figure*}
    \centering
    \begin{minipage}{0.48\textwidth}
        \includegraphics[width=0.7\columnwidth, angle=270]{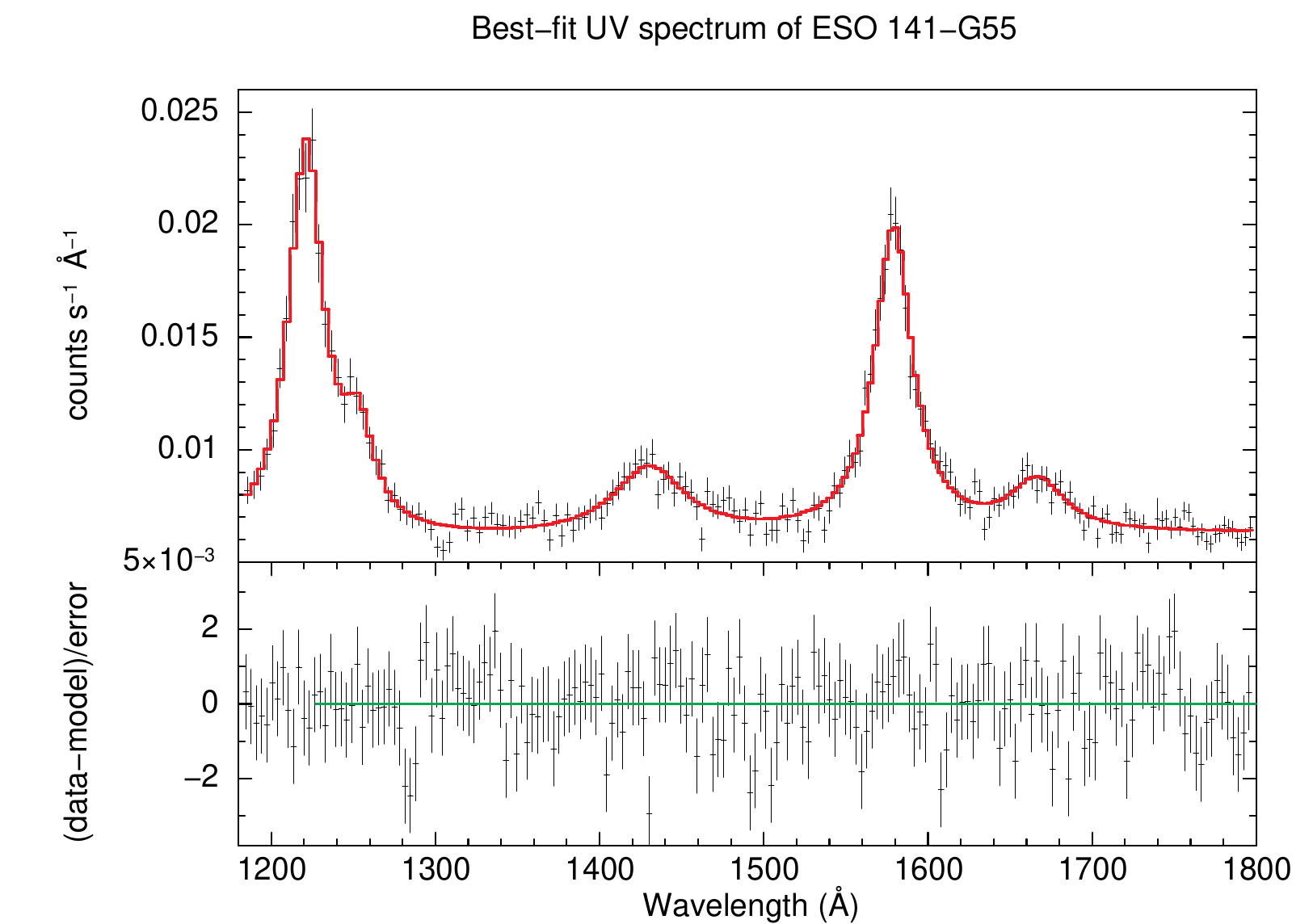}
        \caption{The best-fit SWP UV spectrum (1180-1800~\AA) from IUE is shown (top panel) along with the fitted model (red) consisting of continuum and emission features and the residual of the fitting (bottom panel).}
    \label{Fig2a}
    \end{minipage}
    \hfill
    \begin{minipage}{0.48\textwidth}
        \includegraphics[width=0.7\columnwidth, angle=270]{bestfit_lwr_redshift.eps}
    \caption{The best-fit LWP UV spectrum (2000- 3200~\AA) from IUE is shown (top panel) along with the fitted model (red) consisting of continuum and emission features and the residual of the fitting (bottom panel).}
    \label{Fig2b}
    \end{minipage}
\end{figure*}

To study the flux variability, we model light curves at different wavelengths. The LWR and SWP spectra are modelled separately using the $\chi^2$ minimisation technique in \textsc{XSpec}. To improve the signal-to-noise ratio, we bin the spectra by combining every two adjacent wavelength bins. The size of each bin is 3.35~\AA~after binning. No interpolation or smoothing is used. Both SWP and LWR spectra are fitted, with combinations of a powerlaw function to represent the continuum and multiple Lorentzian components to describe the emission lines, in the wavelength range of 1180-1800~\AA~and 1980-3220~\AA~respectively. The $\chi^2$ per degrees of freedom is found to be 163.49/163 for SWP05514 and 198.47/164 for LWP01708. Using the best fit model, we fit individual IUE spectra by fixing the continuum powerlaw and letting the other parameters vary. The time-averaged UV spectrum along with the model components and residuals for SWP and LWR are shown in Figure \ref{Fig2a} and Figure \ref{Fig2b} respectively, and the corresponding best-fit parameters for SWP are listed in Table \ref{tab2}.
The goodness of fit is quantified by computing the root mean square (RMS) of the residuals, yielding an RMS value of 0.97 for SWP05514 and 1.09 for LWP01708. Hence the deviations between model and data are consistent with the expected observational uncertainties. The powerlaw energy spectral index from the best-fit average UV spectrum is observed to be 2.20 $\pm$ 0.16, which is consistent with that typically observed from AGN \citep{Shull2012, pahari2020evidence}.

Using a convolution model (\textsc{cflux} in \textsc{XSpec}), we derive the continuum-subtracted flux of the peaks of the spectra and their 1$\sigma$ errors. There is one significant peak in LWR, and the SWP show three peaks. Using the same models, we extract the flux of continuum in the range 2098-2221~\AA~for LWR and in the ranges 1763-1783~\AA~and 1356-1376~\AA~for SWP spectra.

\section{lightcurve extraction and results}
\begin{figure*}[htbp]
    \centering
    \includegraphics[width=1.0\textwidth]{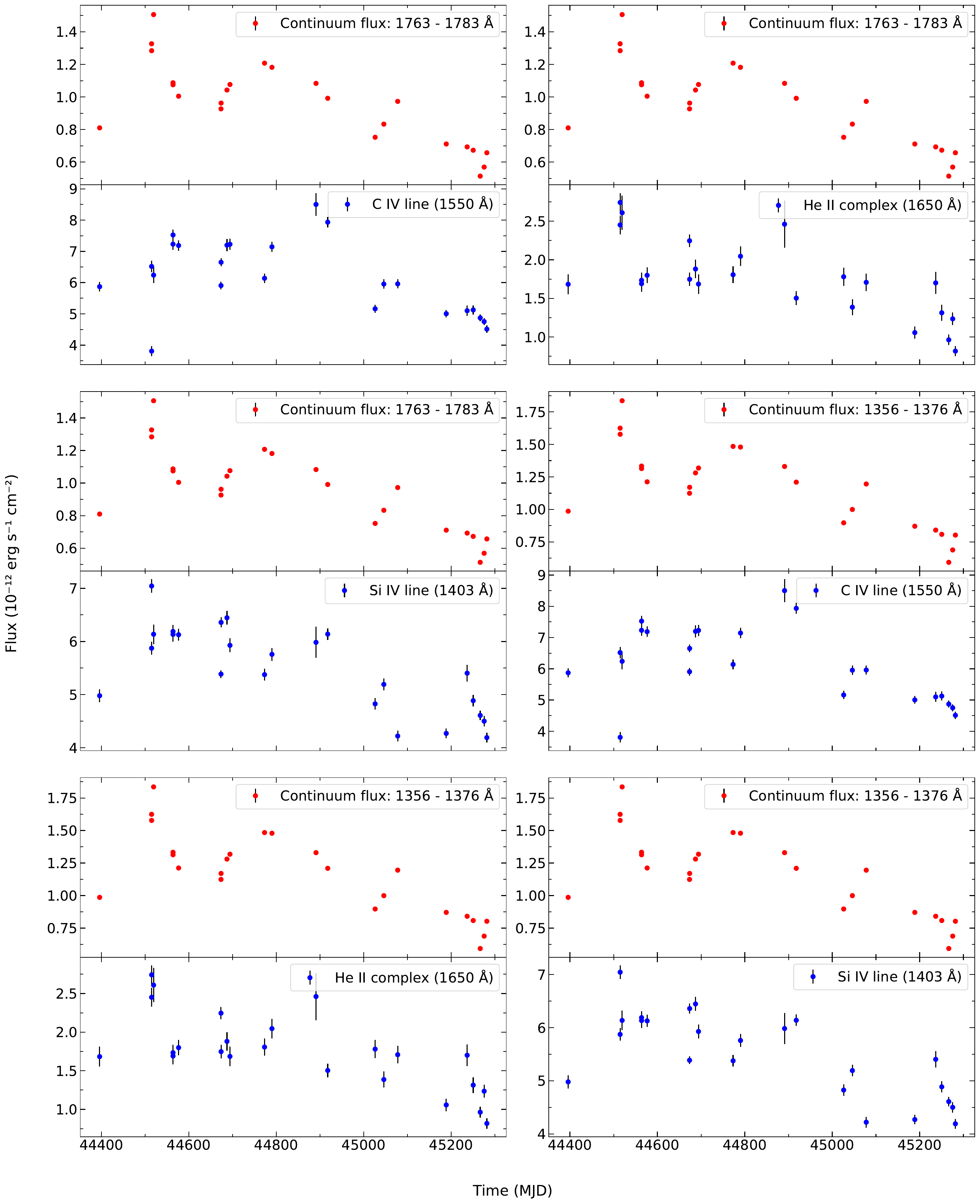}
    \caption{The variation of UV continuum fluxes (in the wavelength range 1356-1376~\AA~and 1763-1783~\AA) for different observations as calculated from fitted UV spectra along with line fluxes for the same observations due to C\textsc{iv} (1550~\AA), He\textsc{ii} (1650~\AA) and Si\textsc{iv} (1403~\AA) lines.}
    \label{Fig3}
\end{figure*}

\begin{figure}
    \centering
    \includegraphics[scale=0.5]{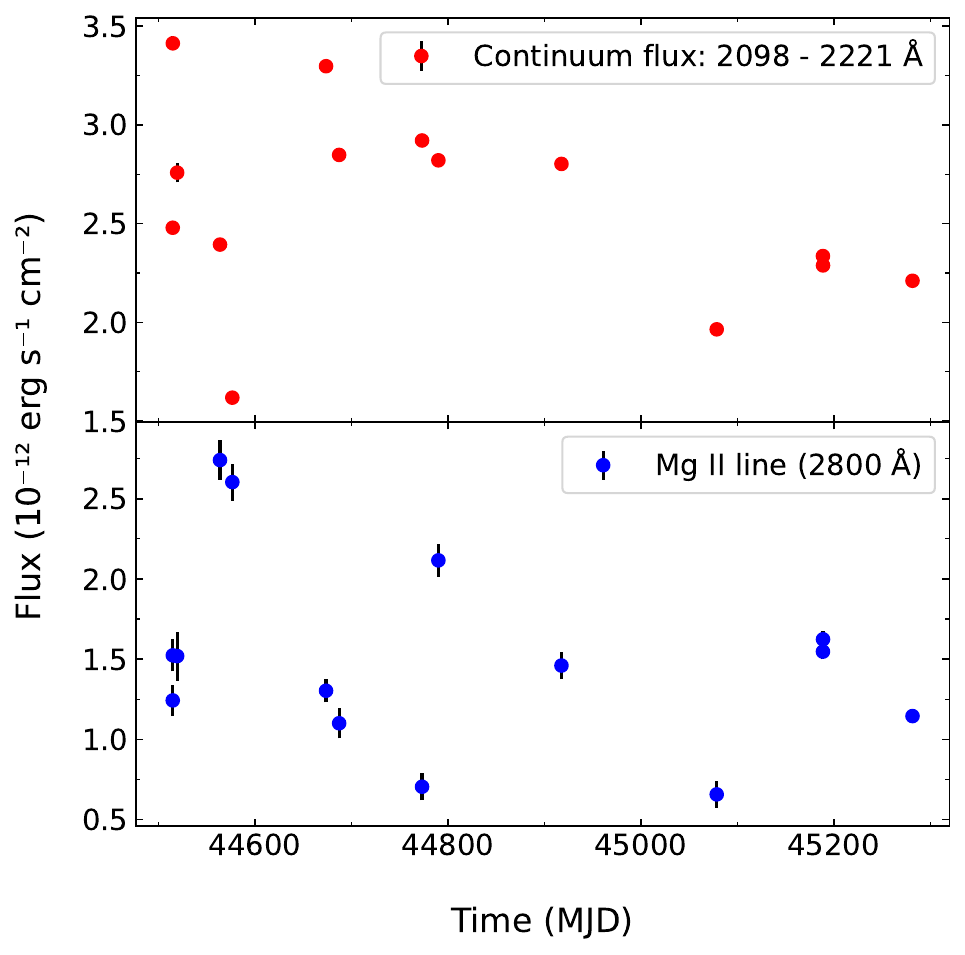}
    \caption{Top panel shows the variation of UV continuum fluxes (in the wavelength range 2098$-$2221~\AA) for different observations as calculated from fitted UV spectra. Bottom panel shows the line flux for the same observations due to the Mg \textsc{ii} line at 2800~\AA.\label{Fig4}}
\end{figure}

The ultraviolet light curves of ESO 141-G55, constructed from both the continuum and line fluxes of the spectral observations as a function of time, are presented in Figures \ref{Fig3} and \ref{Fig4}. The panels display the temporal variations of both the continuum and the principal emission lines. The continuum fluxes are measured in relatively line-free regions at 1763-1783~\AA~(CONT1), 1356-1376~\AA~(CONT2) and 2098-2221~\AA~(CONT3), while the line fluxes correspond to the C\textsc{iv} (1550~\AA), He\textsc{ii} (1650~\AA), Si\textsc{iv} (1403~\AA) and Mg\textsc{ii} (2800~\AA) emission features. All fluxes are plotted as a function of Modified Julian Date (MJD) and expressed in units of $10^{-12}$ erg s$^{-1}$ cm$^{-2}$. The light curves span a total duration of approximately 800 days, with an average cadence (although non-uniform) of $\sim$40 days between successive IUE exposures. The flux values are presented in Table \ref{tab3} and Table \ref{tab4}. In Table \ref{tab5}, we provide the average flux of each continuum and line lightcurves along with their standard deviations (SD). Large SD values for all lightcurves imply the presence of significant variability in both UV line and continuum. The continuum exhibits moderate variability over the monitoring period, with flux changes of about 20–40 \%, whereas the emission lines show stronger fluctuations, in particular the C\textsc{iv} and Si\textsc{iv} lines, which vary by factors of up to two. There is a correlated trend between the continuum and line lightcurves and to quantify it we have measured the time delay between them.

\begin{table}
\centering
\caption{Fitted parameters for SWP05514.\label{tab2}}
\setlength{\tabcolsep}{6pt}
\normalsize
\begin{tabular}{ccll}
\hline
Emission line & Component & Parameter & Value\\
\hline
C\textsc{iv} & lorentz  & LineE (eV) & $7.73 \pm 0.02$\\
     &    & Width (eV) & $0.13 \pm 0.01$\\
      &   & norm          & $0.63 \pm 0.15$\\
Si\textsc{iv} & lorentz  & LineE (eV) & $8.53 \pm 0.05$\\
      &   & Width (eV) & $0.31 \pm 0.01$\\
      &   & norm          & $0.23 \pm 0.11$\\
Ly$\alpha$ & lorentz  & LineE (eV) & $9.85 \pm 0.01$\\
      &   & Width (eV) & $0.17 \pm 0.02$\\
       &  & norm          & $0.58 \pm 0.06$\\
He\textsc{ii} & lorentz  & LineE (eV) & $7.27 \pm 0.01$\\
       &  & Width (eV) & $0.18 \pm 0.02$\\
      &   & norm          & $0.14 \pm 0.01$\\
N\textsc{v} & lorentz  & LineE (eV) & $9.64 \pm 0.09$\\
      &   & Width (eV) & $0.14 \pm 0.02$\\
      &   & norm                 & $0.12 \pm 0.11$\\         
& powerlaw & PhoIndex             & $2.20 \pm 0.16$\\
      &   & norm                 & $0.03 \pm 0.01$\\
\hline
         & & $\chi^2$/dof         & $163.49/163$\\
\hline
\end{tabular}
\end{table}

\begin{table*}[htbp]
\centering
\caption{Fluxes in the unit of $10^{-12}\ \mathrm{erg\ s^{-1}\ cm^{-2}\ }$ are provided from SWP spectral modelling. CONT1 and CONT2 are continuum fluxes calculated in the wavelength ranges of 1763-1783~\AA~ and 1356-1376~\AA~ respectively.\label{tab3}}
%\setlength{\tabcolsep}{6pt}
%\normalsize
\begin{tabular}{cccccc}
\hline
Data ID & CONT1 & CONT2 & C \textsc{iv} & He \textsc{ii}/O complex & Si \textsc{iv} \\
 & flux & flux & flux & flux & flux \\
 \hline
SWP05514 & $1.11\pm 0.00$ & $1.35\pm 0.01$ & $6.60\pm 0.15$ & $2.54\pm 0.13$ & $6.24\pm 0.12$ \\
SWP09200 & $0.81\pm 0.01$ & $0.99\pm 0.01$ & $5.87\pm 0.15$ & $1.68\pm 0.13$ & $4.98\pm 0.12$ \\
SWP10269 & $1.33\pm 0.01$ & $1.62\pm 0.01$ & $6.52\pm 0.19$ & $2.45\pm 0.12$ & $5.87\pm 0.12$ \\
SWP10270 & $1.28\pm 0.01$ & $1.58\pm 0.01$ & $3.81\pm 0.17$ & $2.74\pm 0.12$ & $7.04\pm 0.13$ \\
SWP10298 & $1.51\pm 0.01$ & $1.84\pm 0.01$ & $6.24\pm 0.26$ & $2.61\pm 0.22$ & $6.14\pm 0.18$ \\
SWP10643 & $1.09\pm 0.01$ & $1.33\pm 0.01$ & $7.23\pm 0.18$ & $1.73\pm 0.10$ & $6.19\pm 0.12$ \\
SWP10644 & $1.07\pm 0.01$ & $1.31\pm 0.01$ & $7.52\pm 0.17$ & $1.69\pm 0.11$ & $6.13\pm 0.14$ \\
SWP10741 & $1.00\pm 0.01$ & $1.21\pm 0.01$ & $7.19\pm 0.17$ & $1.80\pm 0.10$ & $6.13\pm 0.11$ \\
SWP13453 & $0.93\pm 0.01$ & $1.12\pm 0.01$ & $5.91\pm 0.13$ & $2.25\pm 0.08$ & $5.39\pm 0.07$ \\
SWP13454 & $0.96\pm 0.01$ & $1.17\pm 0.01$ & $6.65\pm 0.12$ & $1.75\pm 0.09$ & $6.36\pm 0.10$ \\
SWP13563 & $1.04\pm 0.01$ & $1.28\pm 0.01$ & $7.20\pm 0.19$ & $1.88\pm 0.12$ & $6.45\pm 0.13$ \\
SWP13617 & $1.08\pm 0.01$ & $1.32\pm 0.01$ & $7.23\pm 0.18$ & $1.69\pm 0.13$ & $5.93\pm 0.13$ \\
SWP14275 & $1.21\pm 0.01$ & $1.48\pm 0.01$ & $6.14\pm 0.15$ & $1.81\pm 0.11$ & $5.38\pm 0.11$ \\
SWP14399 & $1.18\pm 0.01$ & $1.48\pm 0.01$ & $7.14\pm 0.17$ & $2.05\pm 0.13$ & $5.76\pm 0.12$ \\
SWP15256 & $1.08\pm 0.01$ & $1.33\pm 0.02$ & $8.50\pm 0.37$ & $2.46\pm 0.31$ & $5.98\pm 0.29$ \\
SWP15468 & $0.99\pm 0.01$ & $1.21\pm 0.01$ & $7.93\pm 0.17$ & $1.50\pm 0.09$ & $6.14\pm 0.11$ \\
SWP16428 & $0.75\pm 0.01$ & $0.90\pm 0.01$ & $5.16\pm 0.13$ & $1.78\pm 0.12$ & $4.83\pm 0.11$ \\
SWP16560 & $0.83\pm 0.01$ & $1.00\pm 0.01$ & $5.95\pm 0.15$ & $1.39\pm 0.10$ & $5.19\pm 0.11$ \\
SWP16788 & $0.97\pm 0.01$ & $1.20\pm 0.01$ & $5.96\pm 0.15$ & $1.71\pm 0.11$ & $4.22\pm 0.10$ \\
SWP17622 & $0.71\pm 0.01$ & $0.87\pm 0.01$ & $5.00\pm 0.12$ & $1.06\pm 0.08$ & $4.27\pm 0.09$ \\
SWP18095 & $0.69\pm 0.01$ & $0.84\pm 0.01$ & $5.10\pm 0.16$ & $1.70\pm 0.14$ & $5.40\pm 0.16$ \\
SWP18241 & $0.67\pm 0.01$ & $0.81\pm 0.01$ & $5.13\pm 0.14$ & $1.31\pm 0.10$ & $4.89\pm 0.11$ \\
SWP18383 & $0.51\pm 0.01$ & $0.59\pm 0.01$ & $4.87\pm 0.11$ & $0.96\pm 0.07$ & $4.61\pm 0.09$ \\
SWP18455 & $0.57\pm 0.01$ & $0.69\pm 0.01$ & $4.75\pm 0.11$ & $1.24\pm 0.08$ & $4.50\pm 0.10$ \\
SWP18507 & $0.66\pm 0.01$ & $0.80\pm 0.01$ & $4.51\pm 0.12$ & $0.82\pm 0.07$ & $4.19\pm 0.09$ \\
\hline
\end{tabular}
\end{table*}

\begin{table}
\centering
\caption{Fluxes in the unit of $10^{-12}\ \mathrm{erg\ s^{-1}\ cm^{-2}\ }$ are provided from LWP spectral modelling. CONT3 is the continuum flux calculated in the wavelength ranges of 2098-2221~\AA.\label{tab4}}
\begin{tabular}{ccc}
\hline
Data ID & CONT3 & Mg \textsc{ii}\\
 & flux & flux\\
\hline
LWR04779 & $2.31\pm 0.03$ & $1.64\pm 0.10$\\
LWR08940 & $2.48\pm 0.03$ & $1.24\pm 0.10$\\
LWR08941 & $3.41\pm 0.04$ & $1.52\pm 0.10$\\
LWR08962 & $2.76\pm 0.05$ & $1.52\pm 0.15$\\
LWR09352 & $2.39\pm 0.03$ & $1.74\pm 0.12$\\
LWR09425 & $1.62\pm 0.03$ & $2.60\pm 0.12$\\
LWR10116 & $3.30\pm 0.03$ & $1.30\pm 0.07$\\
LWR10198 & $2.85\pm 0.03$ & $1.10\pm 0.09$\\
LWR10890 & $2.92\pm 0.03$ & $0.71\pm 0.08$ \\
LWR11010 & $2.82\pm 0.03$ & $2.12\pm 0.10$ \\
LWR11952 & $2.80\pm 0.03$ & $1.46\pm 0.08$ \\
LWR13043 & $1.96\pm 0.03$ & $0.66\pm 0.08$ \\
LWP01631 & $2.34\pm 0.02$ & $1.55\pm 0.05$ \\
LWP01632 & $2.29\pm 0.02$ & $1.62\pm 0.05$ \\
LWP01708 & $2.21\pm 0.02$ & $1.15\pm 0.04$ \\
\hline
\end{tabular}
\end{table}

\begin{table}[htbp]
\centering
\caption{Weighted mean flux (in unit of $10^{-12}\ \mathrm{erg\ s^{-1}\ cm^{-2}\ }$) and standard deviation of continuum and emission lines.\label{tab5}}
\begin{tabular}{ccc}
\hline
Lightcurve type & Mean Flux & Standard Deviation\\
\hline
CONT1 & $0.91\pm 0.001$ & $24\pm 04~ \%$ \\
CONT2 & $1.10\pm 0.001$ & $31\pm 05~\%$ \\
CONT3 & $2.46\pm 0.006$ & $42\pm 10~\%$ \\
C \textsc{iv} & $5.82\pm 0.03$ & $107\pm 17~\%$ \\
He \textsc{ii} & $1.56\pm 0.02$ & $49\pm 08~\%$ \\
Si \textsc{iv} & $5.33\pm 0.02$ & $79\pm 13~\%$ \\
Mg \textsc{ii} & $1.39\pm 0.02$ & $43\pm 11~\%$ \\
\hline
\end{tabular}
\end{table}

\begin{table}[htbp]
\centering
\caption{Estimated Time lag (in days) between UV continuum and UV line variabilities from ESO 141--G55 using damped random walk lightcurve modelling. Errors are calculated using MCMC with 68\% confidence.\label{tab6}}
\begin{tabular}{ccc}
\hline
UV Continuum & UV Line & Lag\\
 (wavelength ~\AA)& (wavelength~\AA) & (days)\\
\hline
1356-1376~\AA~ & 1403~\AA~ & 2.92$^{+0.54}_{-0.61}$ \\
1356-1376~\AA~ & 1550~\AA~ & 4.41$^{+0.44}_{-0.54}$ \\
1356-1376~\AA~ & 1650~\AA~ & 4.11$^{+0.35}_{-0.81}$ \\
1763-1783~\AA~ & 1403~\AA~ & 2.78$^{+0.53}_{-0.52}$ \\
1763-1783~\AA~ & 1550~\AA~ & 4.48$^{+0.45}_{-0.58}$ \\
1763-1783~\AA~ & 1650~\AA~ & 4.16$^{+0.39}_{-0.46}$ \\
\hline
\end{tabular}
\end{table}

\subsection{Time lag analysis}\label{sec4}
To measure the reverberation delay between UV continuum and UV lines, we have used \textsc{JAVELIN}. \textsc{JAVELIN} \citep{2011ApJ...735...80Z, 2013ApJ...765..106Z, 2016ApJ...819..122Z} is a widely used tool for estimating time lags between two light curves. It relies on the damped random walk (DRW) model \citep{2009ApJ...698..895K}, which characterises light curve variability using a set of parameters including the time lag to simultaneously fit both curves. The parameter distributions are then sampled using the Markov Chain Monte Carlo (MCMC) method, allowing for a probabilistic understanding of the lag and other model parameters. This methodology has been extensively adopted in earlier RM studies.

\begin{figure}
    \centering
    \includegraphics[width=0.99\columnwidth]{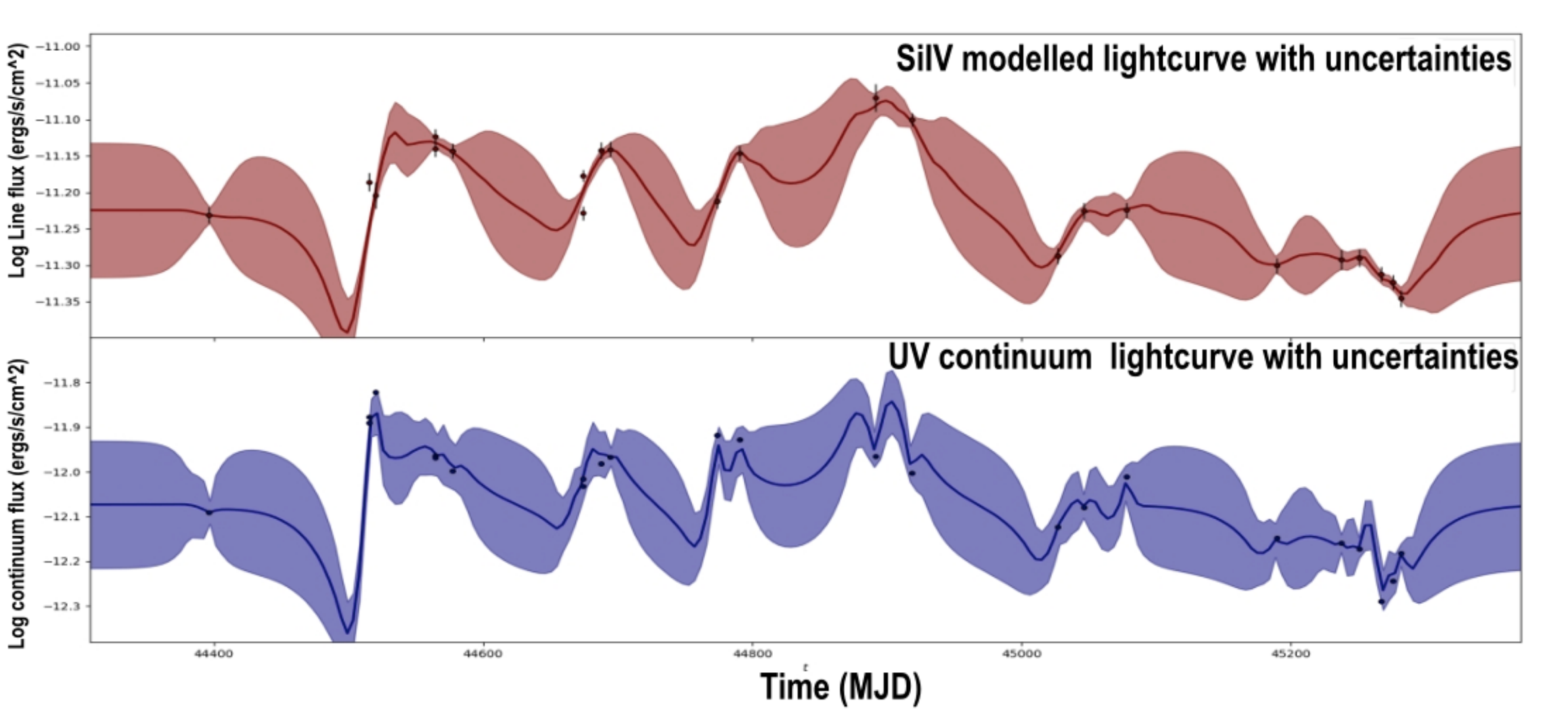}
    \includegraphics[width=0.99\columnwidth]{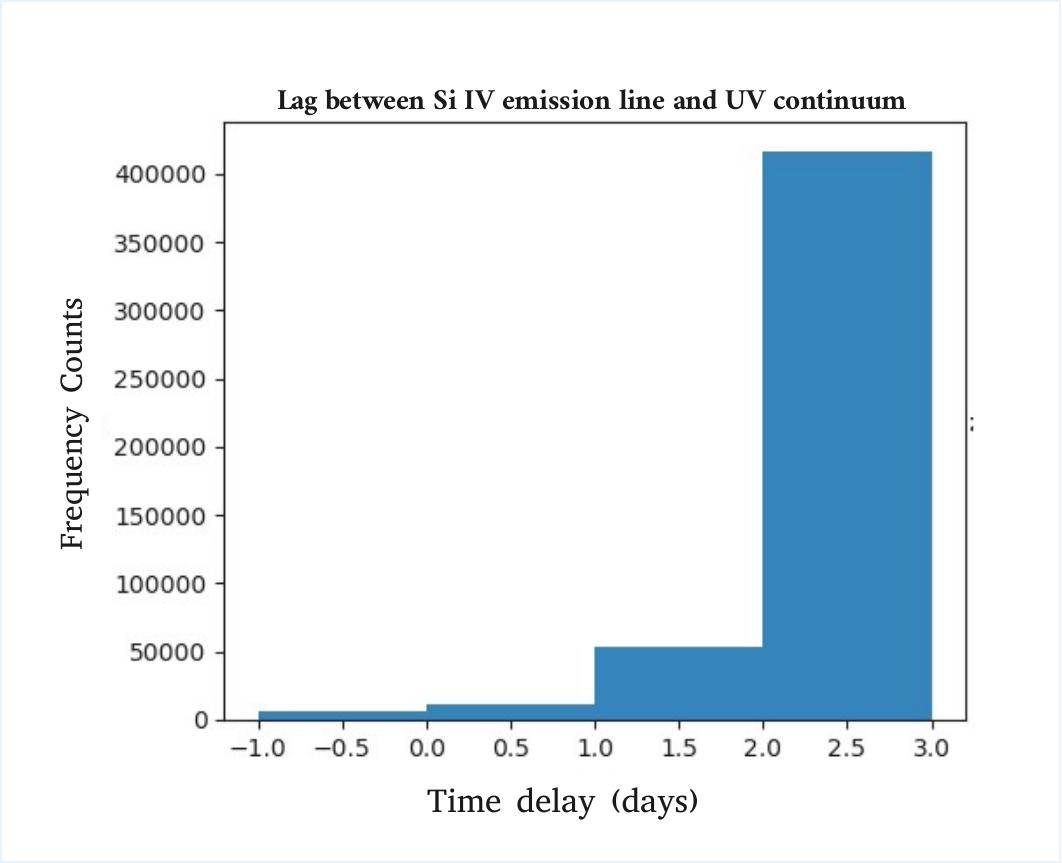}
    \includegraphics[width=0.9\columnwidth]{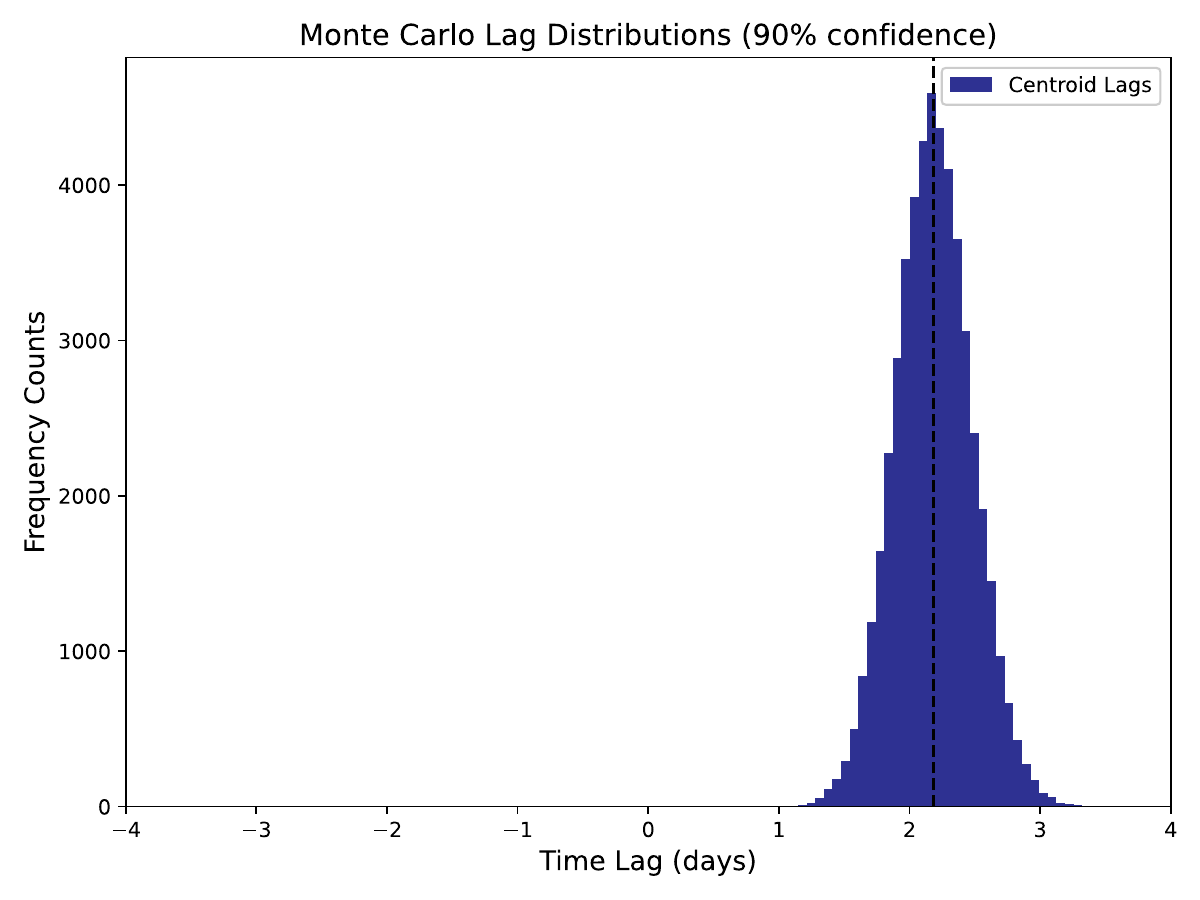}
    \caption{Top panel shows Lightcurves due to SiIV line fluxes (top) and UV continuum fluxes at 1356-1376~\AA~ (bottom), along with fitted \textsc{Pmap} reverberation model and 68\% uncertainty. The middle panel shows the posterior distribution of time delay between the above two lightcurves derived through MCMC simulations. Bottom panel shows the distribution of time delays estimated by ICCF method. Uncertainties on ICCF time delays were estimated using the Flux Randomization/Random Subset Selection (FR/RSS) method with 10000 MC realisation.}\label{Fig5}
\end{figure}

In this work, we utilise JAVELIN’s \textsc{Pmap} model. \textsc{Pmap} assumes that the line-band light curve consists of a combination of the continuum signal and a time-delayed emission line component. The model fitting process yields the relative amplitudes of these two components, offering an estimate of the emission line's contribution within the broadband flux. We have used all continuum and line lightcurves for lag analysis as shown in Figure \ref{Fig3}. However, due to an insufficient number of data points in lightcurves extracted from LWR spectra (shown in Figure \ref{Fig4}), we avoided time lag analysis for LWR data. The model lightcurve flux, along with the uncertainty, is shown in the top panel of Figure \ref{Fig5}. We run MCMC for 100000 chains with 500 walkers after burning the first 10000 steps. A frequency distribution of the time delay between the UV continuum and UV line is shown in the middle panel of Figure \ref{Fig5}. Lag estimations between two continuum bands and three main UV lines: Si\textsc{iv}, C\textsc{iv} and He\textsc{ii} are provided with 68\% errors in Table \ref{tab6}.

In addition to the JAVELIN modeling approach, we also performed an independent cross-correlation analysis using the Interpolated Cross-Correlation Function (ICCF) method \citep{Gaskell1987,Peterson1998}, which is a widely adopted techniques for reverberation lag estimation. The ICCF provides a non-parametric means of estimating time delays by interpolating between unevenly sampled light curves and computing the cross-correlation coefficient as a function of lag. To calculate uncertainties on ICCF lag, we used Monte-Carlo based Flux Randomization(FR) and Random Subset Selection method (RSS) as outlined by \citet{Peterson1998,Peterson2004}. Using 10000 MC realisations, we calculate centroid lag distributions for which cross correlation coefficient is 0.8 or higher. The lag distribution is shown in the bottom panel of Figure \ref{Fig5}. The ICCF lag with 1$\sigma$ error is found to be 2.31 $\pm$0.83 days, which is consistent with the \textsc{JAVELIN} estimations (shown in the middle panel of Figure \ref{Fig5}).

As an independent cross-check on the ICCF and JAVELIN results, we also estimated the time delay using the $z$-transformed discrete correlation function (ZDCF; \citep{alex97}, which is designed for unevenly sampled light curves and employs equal-population binning to mitigate biases arising from sparse or irregular cadence. The ZDCF was computed using data-driven lag bins containing a minimum number of pairs, and uncertainties were estimated via a flux-randomisation and random-subset-selection (FR/RSS) Monte Carlo procedure similar to that adopted in ICCF analyses. From the resulting ZDCF correlation function, the lag was determined from the peak of the correlation distribution. This analysis yields a time delay of  $3.57\pm1.88$~days between the 1356--1376~\AA\ UV continuum and the 1550~\AA\ C\,\textsc{iv}~ emission-line light curve, in good agreement with the corresponding lag reported in Table \ref{tab6}. While this delay lies near the practical detectability limit imposed by the sampling, its consistent recovery across the ZDCF, ICCF, and JAVELIN analyses suggests a short reverberation timescale, although its proximity to the cadence warrants cautious interpretation.

The limitations imposed by the temporal sampling are particularly relevant for the lag measurements obtained using both the ICCF and JAVELIN analyses presented in Section \ref{sec4}. In the ICCF method, the effective sensitivity to a given lag is governed by the distribution of time separations between paired continuum and line measurements that enter the flux-randomization and random subset selection (FR/RSS) realizations, rather than by the median cadence alone \citep{Peterson1998,Peterson2004}. Although the median sampling interval of the IUE light curves is $\Delta t_{\rm med} \sim 10-15$~days, the presence of several clusters of closely spaced observations (Table \ref{tab1}) provides a limited number of correlation pairs at delays of a few days, allowing the ICCF centroid distribution (bottom panel of Figure \ref{Fig2b}) to weakly probe short lags. In the JAVELIN analysis, the continuum variability is modeled as a damped random walk and the emission-line light curve is forward-modeled using the PMAP formalism; while this approach can interpolate between unevenly sampled epochs, its ability to constrain lags significantly shorter than $\Delta t_{\rm med}$ remains dependent on the availability of short-spacing data and on the assumed stochastic variability model \citep{2011ApJ...735...80Z}. Consequently, lag estimatTo calculate uncertainties on ICCF lag, we used Monte-Carlo-based Flux Randomisation (FR) and Random Subset Selection method (RSS) as outlined by \citet{Peterson1998,Peterson2004}.eater model dependence than longer delays. We accordingly interpret these short-lag measurements as tentative, while noting their consistency between the two independent methods.

\section{DISCUSSION}\label{sec5}
We have analysed IUE ultraviolet spectra for ESO 141--G55, which shows substantial variability in both UV continuum and emission lines. Our time lag analysis of the emission lines with respect to the continuum suggests that the lines are possibly emitting from the inner part of BLR which is close to the accretion disk. The following subsection discusses the evidence supporting this hypothesis.
\subsection{Origin of UV line emissions}
Our reverberation lag analysis suggests that UV lines like C \textsc{iv} at 1550~\AA~are delayed to the UV continuum emission by $\sim 4.41$ light days. Such a time scale roughly corresponds to a length scale of 0.004 pc or 4 mpc. Therefore, UV lines may be reprocessed from the outer accretion disk.

To explore further, we have calculated line velocities. From Table \ref{tab2}, using best-fit spectra, the FWHM of C\textsc{iv} line at $7.73\pm 0.02$ eV is $0.126\pm 0.003$ eV. This corresponds to the line width of 26.5~\AA. The velocity width, due to Doppler broadening, will be $\Delta v = \frac{\Delta\lambda}{\lambda}.  c = 4965\pm 311$ km/s. Such a broadening is consistent with the BLR origin of UV lines as suggested by \citet{2021MNRAS.502.2140M}. When similar exercise is applied by the Si\textsc{iv} line, using the best-fit values from Table \ref{tab2}, we get line width of $0.314\pm 0.013$ eV for the line energy $8.53\pm 0.05$ eV. This corresponds to the velocity width of $11031\pm 677$ km/s.
The large velocity width of the Si\textsc{iv} component is naturally consistent with emission from the inner BLR region. Si\textsc{iv} is a high-ionization line and is expected to arise at smaller radii than low-ionization species, where the gravitational potential of the central supermassive black hole dominates the gas dynamics. For ESO~141-G55, a well-studied unobscured Seyfert~1 galaxy with a strong UV/X-ray ionizing continuum, such velocities maybe compatible with virialized gas located at light-day scales. The absence of strong obscuration and the known presence of a classical BLR in this source favor an origin in the inner BLR, where high-ionization species can attain large Doppler widths through Keplerian motion.
An alternative interpretation is that the broad Si\textsc{iv} component traces a radiatively driven disk wind launched from the inner accretion disk. High-ionization UV lines such as Si\textsc{iv} are particularly sensitive to outflowing gas exposed to the intense EUV and soft X-ray radiation field. In ESO~141-G55, which shows evidence for ionized outflows in X-ray spectra with a speed of $\sim$0.17c \citep{porquet2024revealing}, the broad velocity width can be explained by a combination of rotational motion at the wind base and radial acceleration along the flow. Such disk winds are expected to originate at radii comparable to or interior to the classical BLR, naturally producing line widths of order $10^{4}$~km~s$^{-1}$. 
This hypothesis is consistent with our UV line lag measurements. Si\textsc{iv} is delayed by $2.92^{+ 0.54}_{-0.61}$ days with respect to UV continuum, which is significantly shorter than C\textsc{iv} line lag, which is $4.41^{+0.44}_{-0.54}$ days. Therefore, measured lag suggests that the observed Si\textsc{iv} emission probes the dynamical interface between the inner BLR and an accretion-disk-driven outflow.

From the best-fit spectrum, we have computed 1450~\AA~flux (by integrating the flux in the range of 1440$-$1460~\AA) and average fluxes for all observations provided in Table 1. The average 1450~\AA~flux is found to be 1.19 $\pm$ 0.03 $\times$ 10$^{-12}$ ergs s$^{-1}$ cm$^{-2}$. Using the luminosity distance of 163.6 Mpc to ESO 141$-$G55 \citep{2006PASP..118.1711W, planck2020planck}), we have computed the continuum UV luminosity at 1450~\AA~which is 3.9 $\pm$ 0.11 $\times$ 10$^{42}$ ergs s$^{-1}$. Comparing the 1450~\AA~luminosity with the scaling relation from \citet{kaspi2005relationship}, we noted that the BLR size is approximately 4.5 days. This is consistent with the lag between UV lines and continuum measured directly by cross-correlating UV continuum and line lightcurves in our work.  Further multi-wavelength studies of the source will be potentially useful to understand the BLR structure close to the accretion disk.

\section{Conclusions}\label{sec6}
%\backmatter
A detailed ultraviolet spectroscopic investigation of the Seyfert 1 galaxy ESO 141-G55 is carried out using archival data from the International Ultraviolet Explorer (IUE) obtained over multiple epochs between 1978 and 1982. The analysis reveals prominent UV emission lines, including Si\textsc{iv}, C\textsc{iv}, and He\textsc{ii}, arising from ionised gas in the broad-line region (BLR). These features exhibit significant variability and are found to lag behind the UV continuum emission by 2.92, 4.41, and 4.11 light days, respectively. These values correspond to a distance of $\sim$0.004 pc, suggesting that the UV line-emitting regions are located in the inner parts of the BLR, close to the accretion disk. The observed line widths imply high-velocity motions, consistent with Doppler broadening expected from dynamic BLR clouds. The UV continuum luminosity at 1450~\AA, combined with empirical size-luminosity scaling relations, further supports a BLR size of $\sim$4.5 light days, which is in agreement with our lag measurements. For a better understanding of the structure and physical conditions of this region, additional high-cadence, multiwavelength observations will be essential.

\section*{Acknowledgments}
We thank the referee for constructive comments which help to improve the manuscript. MP acknowledges the research facility used at the Department of Physics, IIT Hyderabad.

%\subsection*{Author contributions}

%\subsection*{Financial disclosure}

%None reported.

%\subsection*{Conflict of interest}

%The authors declare no potential conflict of interests.

%\section*{Supporting information}

%\appendix

%\section{Section title of first appendix\label{app1}}

%\nocite{*}% Stoall bib entries - both cited and uncited; comment this line to view only cited bib entries;
\bibliography{Wiley-ASNA}%

@article{Blandford1982,
  author = {Blandford, R. D. and McKee, C. F.},
  title = {Reverberation mapping of the emission line regions of Seyfert galaxies and quasars},
  journal = {ApJ},
  year = {1982},
  volume = {255},
  pages = {419--439}
}

@INPROCEEDINGS{alex97,
       author = {{Alexander}, Tal},
        title = "{Is AGN Variability Correlated with Other AGN Properties? ZDCF Analysis of Small Samples of Sparse Light Curves}",
    booktitle = {Astronomical Time Series},
         year = 1997,
       editor = {{Maoz}, D. and {Sternberg}, A. and {Leibowitz}, E.~M.},
       series = {Astrophysics and Space Science Library},
       volume = {218},
        month = jan,
        pages = {163},
          doi = {10.1007/978-94-015-8941-3_14},
       adsurl = {https://ui.adsabs.harvard.edu/abs/1997ASSL..218..163A}
}

@article{Peterson1993,
  author = {Peterson, B. M.},
  title = {Reverberation mapping of active galactic nuclei},
  journal = {PASP},
  year = {1993},
  volume = {105},
  pages = {247}
}

@ARTICLE{Gaskell1987,
       author = {{Gaskell}, C. Martin and {Peterson}, Bradley M.},
        title = "{The Accuracy of Cross-Correlation Estimates of Quasar Emission-Line Region Sizes}",
      journal = {\apjs},
         year = 1987,
        month = sep,
       volume = {65},
        pages = {1},
          doi = {10.1086/191216},
       adsurl = {https://ui.adsabs.harvard.edu/abs/1987ApJS...65....1G}
}

@ARTICLE{Peterson1998,
       author = {{Peterson}, Bradley M. and {Wanders}, Ignaz and {Horne}, Keith and {Collier}, Stefan and {Alexander}, Tal and {Kaspi}, Shai and {Maoz}, Dan},
        title = "{On Uncertainties in Cross-Correlation Lags and the Reality of Wavelength-dependent Continuum Lags in Active Galactic Nuclei}",
      journal = {\pasp},
         year = 1998,
        month = jun,
       volume = {110},
       number = {748},
        pages = {660-670},
          doi = {10.1086/316177},
archivePrefix = {arXiv},
       eprint = {astro-ph/9802103},
 primaryClass = {astro-ph},
       adsurl = {https://ui.adsabs.harvard.edu/abs/1998PASP..110..660P}
}

@article{Peterson2004,
  author = {Peterson, B. M. and et al.},
  title = {Central Masses and Broad-Line Region Sizes of Active Galactic Nuclei. II. A Homogeneous Analysis of a Large Reverberation-Mapping Database},
  journal = {ApJ},
  year = {2004},
  volume = {613},
  pages = {682}
}

@ARTICLE{calzetti99,
       author = {{Calzetti}, D.},
        title = "{UV Emission and bust properties of high redshift galaxies}",
      journal = {\apss},
         year = 1999,
        month = mar,
       volume = {266},
        pages = {243-253},
archivePrefix = {arXiv},
       eprint = {astro-ph/9902107},
 primaryClass = {astro-ph},
       adsurl = {https://ui.adsabs.harvard.edu/abs/1999Ap&SS.266..243C}
}

@article{Collier1999,
  author = {Collier, S. and et al.},
  title = {Multiwavelength Monitoring of the Seyfert 1 Galaxy NGC 7469. I. Ultraviolet Continuum and Emission-line Variability},
  journal = {MNRAS},
  year = {1999},
  volume = {302},
  pages = {L24}
}

@article{2004NuPhS.132..185G,
  author = {Grupe, D. and Mathur, S.},
  title = {Soft X-ray selected AGN: Are they typical NLS1 galaxies?},
  journal = {Nuclear Physics B - Proceedings Supplements},
  volume = {132},
  pages = {185},
  year = {2004}
}

@article{Elvis1978,
  author = {Elvis, M. and et al.},
  title = {An X-ray Survey of Active Galactic Nuclei},
  journal = {MNRAS},
  year = {1978},
  volume = {183},
  pages = {129}
}

@article{Ward1978,
  author = {Ward, M. and et al.},
  title = {Seyfert galaxies in the Ariel V sky survey},
  journal = {ApJ},
  year = {1978},
  volume = {223},
  pages = {788}
}

@article{1981IAUC.3560....1G,
  author = {Gaskell, C. M. and et al.},
  title = {UV Variability of Seyfert Galaxies},
  journal = {IAU Circ.},
  year = {1981},
  volume = {3560},
  pages = {1}
}

@article{1985ApJ...297..151C,
  author = {Chapman, J. M. and et al.},
  title = {A study of the ultraviolet spectra of Seyfert galaxies},
  journal = {ApJ},
  year = {1985},
  volume = {297},
  pages = {151}
}

@article{Koratkar1991,
  author = {Koratkar, A. and Gaskell, C. M.},
  title = {Structure and Kinematics of the Broad-Line Region in Active Galactic Nuclei},
  journal = {ApJS},
  year = {1991},
  volume = {75},
  pages = {719}
}

@article{Turner1989,
  author = {Turner, T. J. and Pounds, K. A.},
  title = {X-ray spectra of Seyfert galaxies observed with EXOSAT},
  journal = {MNRAS},
  year = {1989},
  volume = {240},
  pages = {833}
}

@article{Turner1991,
  author = {Turner, T. J. and et al.},
  title = {Einstein solid state spectrometer observations of active galactic nuclei},
  journal = {ApJS},
  year = {1991},
  volume = {75},
  pages = {561}
}

@article{Turner1993,
  author = {Turner, T. J. and et al.},
  title = {ROSAT observations of Seyfert galaxies},
  journal = {ApJ},
  year = {1993},
  volume = {407},
  pages = {556}
}

@article{Shakura1973,
  author = {Shakura, N. I. and Sunyaev, R. A.},
  title = {Black holes in binary systems: Observational appearance},
  journal = {A\&A},
  volume = {24},
  pages = {337--355},
  year = {1973}
}

@article{Khachikian1974,
  author = {Khachikian, E. Y. and Weedman, D. W.},
  title = {An atlas of Seyfert galaxies},
  journal = {ApJ},
  volume = {192},
  pages = {581--589},
  year = {1974}
}

@article{Osterbrock1981,
  author = {Osterbrock, D. E.},
  title = {Seyfert galaxies with weak broad H-beta emission lines},
  journal = {ApJ},
  volume = {249},
  pages = {462--470},
  year = {1981}
}

@article{Clavel1991,
  author = {Clavel, J. and others},
  title = {Steps toward Determination of the Size and Structure of the Broad-Line Region in Active Galactic Nuclei. I. an Intensive Hubble Space Telescope, IUE, and Ground-based Study of NGC 5548},
  journal = {ApJ},
  volume = {366},
  pages = {64},
  year = {1991}
}

@article{1978ApJ...223..788W,
  author = {Ward, M. J. and Penston, M. V. and Blades, J. C. and Turtle, A. J.},
  title = {The nature of some X-ray selected Seyfert galaxies},
  journal = {Astrophysical Journal},
  volume = {223},
  pages = {788--796},
  year = {1978}
}

@book{vaucouleurs1991third,
  title={Third Reference Catalogue of Bright Galaxies: Volume II},
  author={Vaucouleurs, G{\'e}rard and Vaucouleurs, Antoinette and Corwin, Harold G and Buta, Ronald J and Paturel, Georges and Fouqu{\'e}, Pascal},
  year={1991},
  publisher={Springer}
}

@article{Cackett2007,
  title={Testing thermal reprocessing in active galactic nuclei accretion discs},
  author={Cackett, Edward M and Horne, Keith and Winkler, Hartmut},
  journal={Monthly Notices of the Royal Astronomical Society},
  volume={380},
  number={2},
  pages={669--682},
  year={2007},
  publisher={The Royal Astronomical Society}
}

@article{Wanders1997,
  title={Steps toward determination of the size and structure of the broad-line region in active galactic nuclei. XI. Intensive monitoring of the ultraviolet spectrum of NGC 7469},
  author={Wanders, I and Peterson, Bradley M and Alloin, D and Ayres, TR and Clavel, J and Crenshaw, D Michael and Horne, K and Kriss, Gerard A and Krolik, Julian Henry and Malkan, Matthew A and others},
  journal={The Astrophysical Journal Supplement Series},
  volume={113},
  number={1},
  pages={69},
  year={1997},
  publisher={IOP Publishing}
}

@article{Ayres1993,
  title={Signal-to-Noise Ratios in IUE Low-Dispersion Spectra. II. Photometrically-Corrected Images},
  author={Ayres, Thomas R},
  journal={Publications of the Astronomical Society of the Pacific},
  volume={105},
  number={687},
  pages={538},
  year={1993},
  publisher={IOP Publishing}
}

@misc{Nichols1993,
  title={IUE NEWSIPS Image Processing System Information Manual: Low Dispersion Data},
  author={Nichols, JS and Garhart, MP and De La Pena, MD and Levay, KL},
  year={1993},
  publisher={Version}
}

@article{Shull2012,
  title={HST-COS observations of AGNs. I. Ultraviolet composite spectra of the ionizing continuum and emission lines},
  author={Shull, J Michael and Stevans, Matthew and Danforth, Charles W},
  journal={The Astrophysical Journal},
  volume={752},
  number={2},
  pages={162},
  year={2012},
  publisher={IOP Publishing}
}

@article{pahari2020evidence,
  title={Evidence for variability time-scale-dependent UV/X-ray delay in Seyfert 1 AGN NGC 7469},
  author={Pahari, Mayukh and McHardy, Ian M and Vincentelli, Federico and Cackett, Edward and Peterson, Bradley M and Goad, Mike and G{\"u}ltekin, Kayhan and Horne, Keith},
  journal={Monthly Notices of the Royal Astronomical Society},
  volume={494},
  number={3},
  pages={4057--4068},
  year={2020},
  publisher={Oxford University Press}
}

@article{porquet2024revealing,
  title={Revealing the burning and soft heart of the bright bare active galactic nucleus ESO 141-G55: X-ray broadband and SED analysis},
  author={Porquet, Delphine and Reeves, James N and Hagen, Scott and Lobban, Andrew and Braito, Valentina and Grosso, Nicolas and Marin, Fr{\'e}d{\'e}ric},
  journal={Astronomy \& Astrophysics},
  volume={689},
  pages={A336},
  year={2024},
  publisher={EDP Sciences}
}

@article{planck2020planck,
  title={Planck 2018 results: VII. Isotropy and statistics of the CMB},
  author={Planck Collaboration and others},
  journal={Astronomy and Astrophysics},
  volume={641},
  pages={A7},
  year={2020},
  publisher={EDP Sciences}
}

@article{kaspi2005relationship,
  title={The relationship between luminosity and broad-line region size in active galactic nuclei},
  author={Kaspi, Shai and Maoz, Dan and Netzer, Hagai and Peterson, Bradley M and Vestergaard, Marianne and Jannuzi, Buell T},
  journal={The Astrophysical Journal},
  volume={629},
  number={1},
  pages={61},
  year={2005},
  publisher={IOP Publishing}
}

@ARTICLE{1988ApJ...333..646E,
       author = {{Edelson}, R.~A. and {Krolik}, J.~H.},
        title = "{The Discrete Correlation Function: A New Method for Analyzing Unevenly Sampled Variability Data}",
      journal = {\apj},
         year = 1988,
        month = oct,
       volume = {333},
        pages = {646},
          doi = {10.1086/166773},
       adsurl = {https://ui.adsabs.harvard.edu/abs/1988ApJ...333..646E}
}

@ARTICLE{2011ApJ...735...80Z,
       author = {{Zu}, Ying and {Kochanek}, C.~S. and {Peterson}, Bradley M.},
        title = "{An Alternative Approach to Measuring Reverberation Lags in Active Galactic Nuclei}",
      journal = {\apj},
         year = 2011,
        month = jul,
       volume = {735},
       number = {2},
          eid = {80},
        pages = {80},
          doi = {10.1088/0004-637X/735/2/80},
archivePrefix = {arXiv},
       eprint = {1008.0641},
 primaryClass = {astro-ph.CO},
       adsurl = {https://ui.adsabs.harvard.edu/abs/2011ApJ...735...80Z}
}

@ARTICLE{2013ApJ...765..106Z,
       author = {{Zu}, Ying and {Kochanek}, C.~S. and {Koz{\l}owski}, Szymon and {Udalski}, Andrzej},
        title = "{Is Quasar Optical Variability a Damped Random Walk?}",
      journal = {\apj},
         year = 2013,
        month = mar,
       volume = {765},
       number = {2},
          eid = {106},
        pages = {106},
          doi = {10.1088/0004-637X/765/2/106},
archivePrefix = {arXiv},
       eprint = {1202.3783},
 primaryClass = {astro-ph.CO},
       adsurl = {https://ui.adsabs.harvard.edu/abs/2013ApJ...765..106Z}
}

@ARTICLE{2016ApJ...819..122Z,
       author = {{Zu}, Ying and {Kochanek}, C.~S. and {Koz{\l}owski}, Szymon and {Peterson}, B.~M.},
        title = "{Application of Stochastic Modeling to Analysis of Photometric Reverberation Mapping Data}",
      journal = {\apj},
         year = 2016,
        month = mar,
       volume = {819},
       number = {2},
          eid = {122},
        pages = {122},
          doi = {10.3847/0004-637X/819/2/122},
archivePrefix = {arXiv},
       eprint = {1310.6774},
 primaryClass = {astro-ph.CO},
       adsurl = {https://ui.adsabs.harvard.edu/abs/2016ApJ...819..122Z}
}

@ARTICLE{2009ApJ...698..895K,
       author = {{Kelly}, Brandon C. and {Bechtold}, Jill and {Siemiginowska}, Aneta},
        title = "{Are the Variations in Quasar Optical Flux Driven by Thermal Fluctuations?}",
      journal = {\apj},
         year = 2009,
        month = jun,
       volume = {698},
       number = {1},
        pages = {895-910},
          doi = {10.1088/0004-637X/698/1/895},
archivePrefix = {arXiv},
       eprint = {0903.5315},
 primaryClass = {astro-ph.CO},
       adsurl = {https://ui.adsabs.harvard.edu/abs/2009ApJ...698..895K}
}

@ARTICLE{2021MNRAS.502.2140M,
       author = {{Mandal}, Amit Kumar and {Rakshit}, Suvendu and {Stalin}, C.~S. and {Petrov}, R.~G. and {Mathew}, Blesson and {Sagar}, Ram},
        title = "{Estimation of the size and structure of the broad line region using Bayesian approach}",
      journal = {\mnras},
         year = 2021,
        month = apr,
       volume = {502},
       number = {2},
        pages = {2140-2157},
          doi = {10.1093/mnras/stab012},
archivePrefix = {arXiv},
       eprint = {2101.00802},
 primaryClass = {astro-ph.GA},
       adsurl = {https://ui.adsabs.harvard.edu/abs/2021MNRAS.502.2140M}
}

@ARTICLE{2006PASP..118.1711W,
       author = {{Wright}, E.~L.},
        title = "{A Cosmology Calculator for the World Wide Web}",
      journal = {\pasp},
         year = 2006,
        month = dec,
       volume = {118},
       number = {850},
        pages = {1711-1715},
          doi = {10.1086/510102},
archivePrefix = {arXiv},
       eprint = {astro-ph/0609593},
 primaryClass = {astro-ph},
       adsurl = {https://ui.adsabs.harvard.edu/abs/2006PASP..118.1711W}
}

@ARTICLE{2016ApJ...821...56F,
       author = {{Fausnaugh}, M.~M. and {Denney}, K.~D. and {Barth}, A.~J. and {Bentz}, M.~C. and {Bottorff}, M.~C. and {Carini}, M.~T. and {Croxall}, K.~V. and {De Rosa}, G. and {Goad}, M.~R. and {Horne}, Keith and {Joner}, M.~D. and {Kaspi}, S. and {Kim}, M. and {Klimanov}, S.~A. and {Kochanek}, C.~S. and {Leonard}, D.~C. and {Netzer}, H. and {Peterson}, B.~M. and {Schn{\"u}lle}, K. and {Sergeev}, S.~G. and {Vestergaard}, M. and {Zheng}, W.-K. and {Zu}, Y. and {Anderson}, M.~D. and {Ar{\'e}valo}, P. and {Bazhaw}, C. and {Borman}, G.~A. and {Boroson}, T.~A. and {Brandt}, W.~N. and {Breeveld}, A.~A. and {Brewer}, B.~J. and {Cackett}, E.~M. and {Crenshaw}, D.~M. and {Dalla Bont{\`a}}, E. and {De Lorenzo-C{\'a}ceres}, A. and {Dietrich}, M. and {Edelson}, R. and {Efimova}, N.~V. and {Ely}, J. and {Evans}, P.~A. and {Filippenko}, A.~V. and {Flatland}, K. and {Gehrels}, N. and {Geier}, S. and {Gelbord}, J.~M. and {Gonzalez}, L. and {Gorjian}, V. and {Grier}, C.~J. and {Grupe}, D. and {Hall}, P.~B. and {Hicks}, S. and {Horenstein}, D. and {Hutchison}, T. and {Im}, M. and {Jensen}, J.~J. and {Jones}, J. and {Kaastra}, J. and {Kelly}, B.~C. and {Kennea}, J.~A. and {Kim}, S.~C. and {Korista}, K.~T. and {Kriss}, G.~A. and {Lee}, J.~C. and {Lira}, P. and {MacInnis}, F. and {Manne-Nicholas}, E.~R. and {Mathur}, S. and {McHardy}, I.~M. and {Montouri}, C. and {Musso}, R. and {Nazarov}, S.~V. and {Norris}, R.~P. and {Nousek}, J.~A. and {Okhmat}, D.~N. and {Pancoast}, A. and {Papadakis}, I. and {Parks}, J.~R. and {Pei}, L. and {Pogge}, R.~W. and {Pott}, J.-U. and {Rafter}, S.~E. and {Rix}, H.-W. and {Saylor}, D.~A. and {Schimoia}, J.~S. and {Siegel}, M. and {Spencer}, M. and {Starkey}, D. and {Sung}, H.-I. and {Teems}, K.~G. and {Treu}, T. and {Turner}, C.~S. and {Uttley}, P. and {Villforth}, C. and {Weiss}, Y. and {Woo}, J.-H. and {Yan}, H. and {Young}, S.},
        title = "{Space Telescope and Optical Reverberation Mapping Project. III. Optical Continuum Emission and Broadband Time Delays in NGC 5548}",
      journal = {\apj},
         year = 2016,
        month = apr,
       volume = {821},
       number = {1},
          eid = {56},
        pages = {56},
          doi = {10.3847/0004-637X/821/1/56},
archivePrefix = {arXiv},
       eprint = {1510.05648},
 primaryClass = {astro-ph.GA},
       adsurl = {https://ui.adsabs.harvard.edu/abs/2016ApJ...821...56F}
}

@ARTICLE{2022ApJ...940...20G,
       author = {{Guo}, Hengxiao and {Barth}, Aaron J. and {Wang}, Shu},
        title = "{Active Galactic Nuclei Continuum Reverberation Mapping Based on Zwicky Transient Facility Light Curves}",
      journal = {\apj},
         year = 2022,
        month = nov,
       volume = {940},
       number = {1},
          eid = {20},
        pages = {20},
          doi = {10.3847/1538-4357/ac96ec},
archivePrefix = {arXiv},
       eprint = {2207.06432},
 primaryClass = {astro-ph.GA},
       adsurl = {https://ui.adsabs.harvard.edu/abs/2022ApJ...940...20G}
}

@ARTICLE{2023ApJ...958..195C,
       author = {{Cackett}, Edward M. and {Gelbord}, Jonathan and {Barth}, Aaron J. and {De Rosa}, Gisella and {Edelson}, Rick and {Goad}, Michael R. and {Homayouni}, Yasaman and {Horne}, Keith and {Kara}, Erin A. and {Kriss}, Gerard A. and {Korista}, Kirk T. and {Landt}, Hermine and {Plesha}, Rachel and {Arav}, Nahum and {Bentz}, Misty C. and {Boizelle}, Benjamin D. and {Dalla Bont{\`a}}, Elena and {Dehghanian}, Maryam and {Donnan}, Fergus and {Du}, Pu and {Ferland}, Gary J. and {Fian}, Carina and {Filippenko}, Alexei V. and {Gonz{\'a}lez Buitrago}, Diego H. and {Grier}, Catherine J. and {Hall}, Patrick B. and {Hu}, Chen and {Ili{\'c}}, Dragana and {Kaastra}, Jelle and {Kaspi}, Shai and {Kochanek}, Christopher S. and {Kova{\v{c}}evi{\'c}}, Andjelka B. and {Kynoch}, Daniel and {Li}, Yan-Rong and {McLane}, Jacob N. and {Mehdipour}, Missagh and {Miller}, Jake A. and {Montano}, John and {Netzer}, Hagai and {Panagiotou}, Christos and {Partington}, Ethan and {{\v{C}}. Popovi{\'c}}, Luka and {Proga}, Daniel and {Rogantini}, Daniele and {Sanmartim}, David and {Siebert}, Matthew R. and {Storchi-Bergmann}, Thaisa and {Vestergaard}, Marianne and {Wang}, Jian-Min and {Waters}, Tim and {Zaidouni}, Fatima},
        title = "{AGN STORM 2. IV. Swift X-Ray and Ultraviolet/Optical Monitoring of Mrk 817}",
      journal = {\apj},
         year = 2023,
        month = dec,
       volume = {958},
       number = {2},
          eid = {195},
        pages = {195},
          doi = {10.3847/1538-4357/acfdac},
archivePrefix = {arXiv},
       eprint = {2306.17663},
 primaryClass = {astro-ph.HE},
       adsurl = {https://ui.adsabs.harvard.edu/abs/2023ApJ...958..195C}
}

% \section*{Author Biography}
% (if applicable)

\end{document}